\documentclass[ aip, jmp, amsmath,amssymb, reprint]{revtex4-1}

\usepackage{graphicx}
\usepackage{epstopdf}
\usepackage{dcolumn}
\usepackage{bm}
\usepackage{color}
\usepackage[normalem]{ulem}

\newcommand{\comment}[1]{}

\begin{document}

\preprint{AIP/123-QED}

\title[Spatial self-organization in hybrid models of multicellular adhesion]{Spatial self-organization in hybrid models of multicellular adhesion}

\author{Bonforti, A.}
\thanks{These two authors contributed equally to this work}
 \affiliation{ICREA - Complex Systems Lab, UPF, Dr Aiguad\'e 88, 08003 Barcelona, Spain}
 \affiliation{Institut de Biologia Evolutiva (CSIC-Universitat Pompeu Fabra), Passeig Mar\'itim de la Barceloneta 37, 08003 Barcelona, Spain.}
 \affiliation{Department of Experimental and Health Sciences, UPF, 08003 Barcelona, Spain}
 \affiliation{CIDI - Sant Joan de Deu Research Foundation, 08950 Barcelona, Spain}

\author{Duran-Nebreda, S.}
\thanks{These two authors contributed equally to this work}
 \affiliation{ICREA - Complex Systems Lab, UPF, Dr Aiguad\'e 88, 08003 Barcelona, Spain}
 \affiliation{Institut de Biologia Evolutiva (CSIC-Universitat Pompeu Fabra), Passeig Mar\'itim de la Barceloneta 37, 08003 Barcelona, Spain.}
  \affiliation{Department of Experimental and Health Sciences, UPF, 08003 Barcelona, Spain}
 
\author{Monta\~nez, R.}
 \affiliation{ICREA - Complex Systems Lab, UPF, Dr Aiguad\'e 88, 08003 Barcelona, Spain}
 \affiliation{Institut de Biologia Evolutiva (CSIC-Universitat Pompeu Fabra), Passeig Mar\'itim de la Barceloneta 37, 08003 Barcelona, Spain.}
  \affiliation{Department of Experimental and Health Sciences, UPF, 08003 Barcelona, Spain}

\author{Sol\'e, R. V.}
\thanks{Corresponding author}
 \affiliation{ICREA - Complex Systems Lab, UPF, Dr Aiguad\'e 88, 08003 Barcelona, Spain}
 \affiliation{Institut de Biologia Evolutiva (CSIC-Universitat Pompeu Fabra), Passeig Mar\'itim de la Barceloneta 37, 08003 Barcelona, Spain.}
 \affiliation{Department of Experimental and Health Sciences, UPF, 08003 Barcelona, Spain}
 \affiliation{Santa Fe Institute, 1399 Hyde Park Road, Santa Fe NM 87501, USA}

\begin{abstract}
Spatial self-organization emerges in distributed systems exhibiting local interactions when nonlinearities and the appropriate propagation 
of signals are at work. These kinds of phenomena can be modeled with different frameworks, typically cellular automata or reaction-diffusion
systems. A different class of dynamical processes involves the correlated movement of agents over space, which can be mediated through 
chemotactic movement or minimization of cell-cell interaction energy. A classic example of the latter is given by the formation 
of spatially segregated assemblies when cells display differential adhesion.
Here we consider a new class of dynamical models, involving 
cell adhesion among two stochastically exchangeable cell states as a minimal model capable of exhibiting well-defined, ordered spatial patterns.
Our results suggest that a whole space of pattern-forming rules is hosted by the combination of physical differential adhesion and the value of probabilities modulating cell phenotypic switching, showing that Turing-like patterns can be obtained without resorting to reaction-diffusion processes.
If the model is expanded allowing cells to proliferate and die in an environment where diffusible nutrient and toxic waste are at play,
different phases are observed, characterized by regularly spaced patterns.
The analysis of the parameter space reveals that certain phases reach higher population levels than other modes of organization.
A detailed exploration of the mean-field theory is also presented.
Finally we let populations of cells with different adhesion matrices compete for reproduction, showing that, in our model, structural organization can improve the fitness of a given cell population.
The implications of these results for ecological and evolutionary models of pattern formation and the emergence of multicellularity are outlined.

\end{abstract}

\keywords{Turing Patterns, Artificial life, Evolutionary Transitions, Cellular Automata, Complexity}

\maketitle

\section{\label{sec:level1}Introduction}

The evolution of life shows an overall trend towards an increase in size and complexity\cite{Bonner1988,Carroll2001}. One of the determining major innovations that have allowed biological systems to achieve a high degree of complexity has been the evolution of multicellularity and the emergence of supra-cellular hierarchies beyond single-cell organization\cite{Nedelcu2015}. Together with multicellularity, mechanisms to maintain stable phenotypes that underly consistent division of labor had to be developed\cite{Bonner2001, Newman1990}. 

The study of the origins of form have a long tradition in biology\cite{Waddington}. Initiated by Turing\cite{Turing1952} and Rashevsky\cite{Rashevsky}, numerous attempts to formalize a mathematical description of pattern formation have been made. As a result, spatial instabilities were proposed as a powerful rationale for the creation of spatial order, out of random fluctuations, around a homogeneous state in reaction-diffusion systems\cite{Murray1981,Sole1,Sole2,TheraulazPNAS,Jilkine2011}. The main feature of reaction-diffusion systems is the presence of diffusion-driven instabilities under certain parametric conditions, by which small perturbations in the system are amplified, leading to ordered spatial patterns. This family of models has been systematically studied \cite{Meinhardt2000, Goodwin1994} and provides the basis for several natural mechanisms of pattern formation \cite{koch, econoumou, raspopovic, lejeune, sick}. The structures generated by these processes have a characteristic scale whose wavelength depends on the model parameters.

Along with this class of pattern-forming mechanisms, another possible class of models capable of organizing structures in space is based on cell-cell differential adhesion\cite{Goel1970,NewmanBaht2008}. Such a mechanism explains the spatial re-arrangement of different cells belonging to disrupted tissues when mixed together {\em in vitro}\cite{ForgacsNewman2005}. After a transient, clusters involving cells of the same class are often observed as spatially segregated from other cell types by means of the formation of well defined boundaries or layers. In this case, the underlying mechanism explaining the origin of patterns is that of energy-minimization dynamics, similar to the one used in physics for strongly interacting particle systems. 
Both reported mechanisms are crucial in the formation of natural self-organized structures in  developing embryos \cite{ForgacsFoty1998, Foty2005}, and have been connected to the early forms of multicellularity \cite{NewmanBaht2009, NewmanBaht2008}.

In this paper we focus our attention on the early stages of the transition towards multicellularity, where the explicit connection between fitness, function and structure has been particularly difficult to elucidate and, thus, is commonly overlooked. To assess whether the structural organization of multicellular assemblies is related to differential fitness, we have developed an embodied computational model where Turing-like structures appear, stemming from differential adhesion and stochastic phenotypic switching. Fitness is intrinsically obtained by the introduction of a limiting nutrient and the production of a toxic waste byproduct, which respectively increase cell reproduction or death. One of the two cellular states is able to process waste at the cost of reduced proliferation. We observe that different parameter sets produce different spatial patterns, and that spatial organization can have a role in increasing fitness. Finally, we discuss the implications of such results for the transition from unicellular to multicellular organisms and for the evolution of complexity.

\begin{figure}[h]
\label{corerules}
\begin{center}
\includegraphics[scale=0.50]{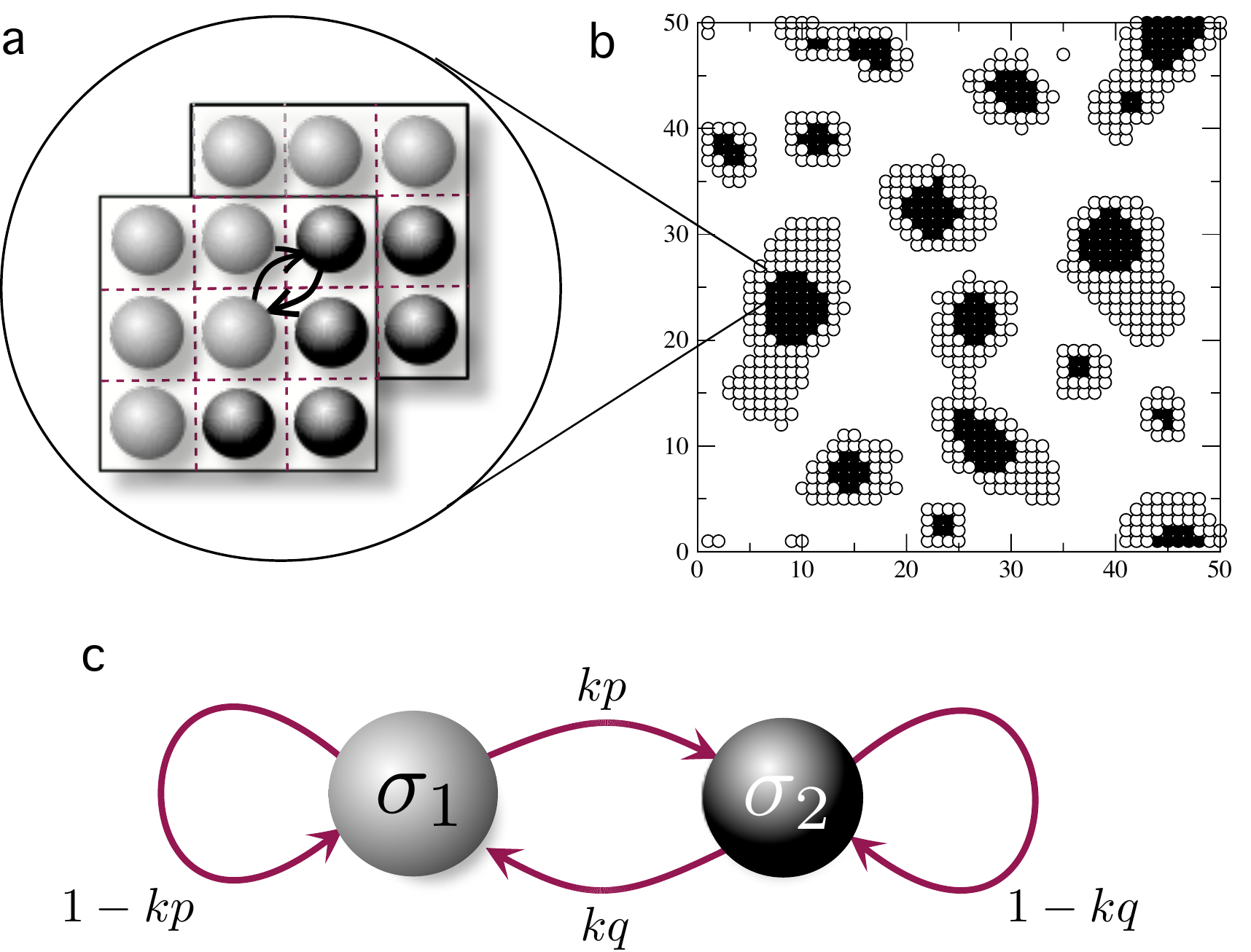}
\end{center}
\caption{(a) Schematic representation of the cell sorting algorithm as described by Steinberg \cite{Steinberg1975}. Cells of different types coexist in a regular square lattice along with empty medium, swapping positions with random neighbouring cells if a particular potential ${\cal H}$ is minimized. (b) The final global configuration is a direct consequence of the micro rules imposed by the adhesion matrix. (c) Markovian process modelling the cell state transitions used in this paper. This very simple approach can aptly describe persister cell dynamics and phase variation phenomena \cite{Darmon2014}.}   
\end{figure}

\section{\label{sec:level1}The model}

Our model considers a population of cells living on a two-dimensional square lattice (Fig. 1), along with empty medium, and following the rules of a cellular Potts model as described by Steinberg \cite{Steinberg1975} (see also\cite{GranerGlazier1992}). Within this framework, cells are discrete entities that occupy single lattice positions, have an associated state ($\sigma_n$) and move across the lattice trying to minimize their energetic potential. Two states correspond to cellular phenotypes, namely white cells ($\sigma_1$) and black cells ($\sigma_2$), while state $\sigma_0$ represents empty space. 

In this paper we build and analyse two different expansions of the basic Potts model: a \emph{hybrid differential adhesion-stochastic phenotipic switching} (DA-SPS) model and an \emph{ecology and competition} (EC) model. 
In the DA-SPS model, cells are sorted by differential adhesion and can reversibly switch their phenotypes. In the EC model we include a simple metabolism by adding nutrient and toxic waste, whose concentrations drive cell proliferation and death. In the following sections we explain how cellular adhesion, phenotypic switching and metabolism are implemented in our models.

\subsection{Differential cellular adhesion}

The cell sorting process fundamentally occurs due to the differences in adhesion energy between states. Following Steinberg's differential adhesion hypothesis (DAH) we assume that the adhesion kinetics are driven by the minimization of adhesion energy between lattice sites, being cells more or less prone to remain together, and avoid or maximize contact with the external medium \cite{ForgacsNewman2005,Foty2005,Steinberg1975,Hogeweg2000}. The strength of interactions among different states can be defined by means of an adhesion matrix ${\bf  {\cal J}}$:
$$
{\bf  {\cal J}} = 
\left( \begin{array}{ccc}
{\cal J}_{(\sigma_0,\sigma_0)} & {\cal J}_{(\sigma_0,\sigma_1)} & {\cal J}_{(\sigma_0,\sigma_2)}\\
{\cal J}_{(\sigma_1,\sigma_0)} & {\cal J}_{(\sigma_1,\sigma_1)} & {\cal J}_{(\sigma_1,\sigma_2)}\\
{\cal J}_{(\sigma_2,\sigma_0)} & {\cal J}_{(\sigma_2,\sigma_1)} & {\cal J}_{(\sigma_2,\sigma_2)} \end{array} \right).
$$ 
Each term $J_{(a,b)}$ in this matrix describes how favourable the pairwise interaction between two states is.
The matrix is symmetric, i.e. ${\cal J}_{(a,b)}={\cal J}_{(b,a)}$, and has ${\cal J}_{(\sigma_0,\sigma_0)}=0$ always. To avoid confusion, we will use the notation $\sigma_n$ when we refer to a given state, and the notation $S_{ij}$  to indicate a state occupying a given lattice site $(i,j)$. 
\begin{figure*}[htpb]
\label{differences}
\begin{center}
\includegraphics[scale=0.718]{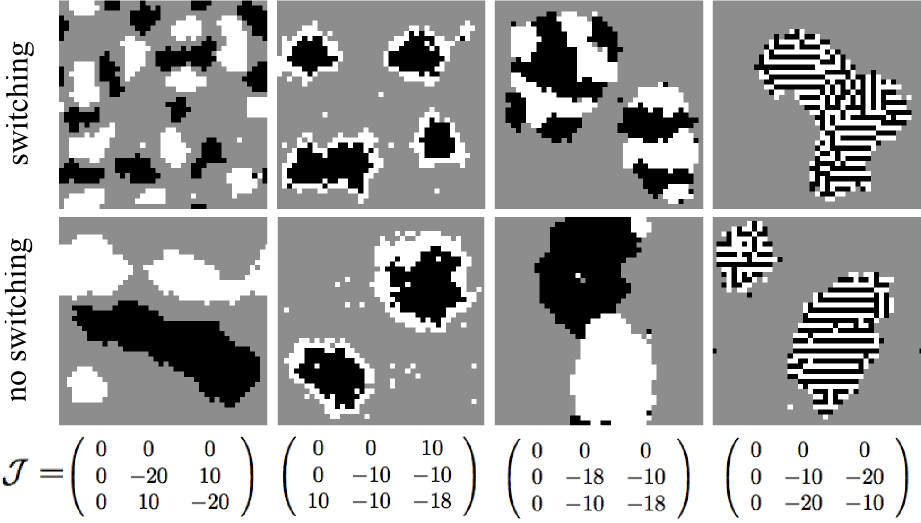}
\end{center}
\caption{Effects of stochastic switching on pattern formation resulting from differential adhesion. In the first row we display the patterns formed after $10^5$ iterations of Boltzmann update rules in a $40\times40$ thoroidal lattice with $0.35$ occupation. The second row displays the resulting patterns of the same conditions when  switching between types is implemented ($p=q=1, \kappa=10^{-3}$). By adding a stochastic switching rule, one can observe changes in the characteristic size of the structures, and also in the spatial arrangement of cell types. Finally, in the third row we show the adhesion matrix for each simulated condition.}   
\end{figure*}
The underlying idea here is that cells will tend to move whenever this allows the system to reach a lower energy. It can be shown that the energy function $\cal H$ in a given position $(i,j)$ can be defined as follows:
\begin{align}
{\cal H}_{ij} = \sum_{S_{kl} \in \Gamma_{ij}}  {\cal J}_{S_{kl},S_{ij}},
\end{align}
where $\Gamma_{ij}$ is the set defined by the eight nearest neighbours of a cell in position $(i,j)$ (Moore's neighbourhood), each of which occupies a position $(k,l)$, and has a defined state $S_{kl}$. To calculate the probability that the cell in ($i,j$) will swap with a randomly chosen neighbour, we calculate the energy function when no swap occurs. This energy function, named ${\cal H^*}$, consists of two terms, one involving the cell in its original position and its neighbourhood set $\Gamma_{ij}$, and another involving the cell's neighbour, located in $i',j'$, with its neighbourhood $\Gamma_{i',j'}$. We then virtually swap the positions of the cell with its neighbour in $i',j'$, and calculate the energy function when swap occurs (${\cal H}'$). The energy difference is then defined as:
\begin{align}
\Delta {\cal H} = {\cal H}' - {\cal H^*}.
\end{align}
When the difference is negative, a decrease in the global energy occurs and the states will swap position. Instead, when $\Delta {\cal H}>0$, the larger the difference the less the swap is likely to happen, with a probability following the Boltzmann distribution. If we indicate $P(S_{ij} \rightarrow S_{kl})$ as the probability that our cell moves from $(i,j)$ to $(k,l)$, it can be shown that:
\begin{align}
P(S_{ij} \rightarrow S_{kl}) = {1 \over 1 + e^{\Delta {\cal H}/T}} , 
\end{align}
where the parameter $T$ is a noise factor acting as a `temperature', essentially tuning the degree of determinism of our system. The Boltzmann factor $e^{\Delta {\cal H}/T}$ acts in such a way that if $\Delta {\cal H}=0$, the probability of swapping is $1/2$. Note that cell-cell (and cell-medium) interactions are local (Fig. 1), meaning that a cell in ($i,j$) interacts only with the set $\Gamma_{ij}$ of its eight nearest neighbours. 

Depending on the form of adhesion matrix, different patterns can be observed in a cell sorting system with two cell types. Unless otherwise indicated, in our simulations we will apply a symmetric adhesion matrix (see Fig. SM-1), where cells tend to attach preferentially to other cells in the same state, and secondarily to cells of the opposite state, while attachment with empty space is not favoured. 

The values of the adhesion matrix determine the structure of patterns formed by the cell sorting algorithm (Fig. 2), which can be perturbed by the effects of phenotypic switching. 

\subsection{Stochastic phenotypic switching}
Cells can perform reversible transitions between their states $\sigma_1$ and $\sigma_2$, similarly to phase variation and persistence in natural bacterial populations \cite{Hallet2001, Balaban2004, Darmon2014}. Switching is regulated by transition probabilities $p$ and $q$,
$$P(\sigma_1 \rightarrow \sigma_2) = \kappa p_{ij} \quad \quad \quad P(\sigma_2 \rightarrow \sigma_1) = \kappa q_{ij}$$
where $\kappa$ is a fixed scaling factor, introduced to regulate the relative speed between adhesion kinetics and phenotypic switching. 
By simply adding SPS to a classical DA model, cell sorting properties can change drastically for some adhesion matrices  (see Fig. 2).
It is worth mentioning that in SPS the transitions between states are not dependent on any molecular cue nor any cellular memory beyond their current state.

\subsection{Metabolism}
Cellular metabolism is defined by two simple pathways: the ability of both $\sigma_1$(white) and $\sigma_2$(black) phenotypes to transform nutrient $N$ (constantly added to the lattice) into cellular energy $E$ and waste byproduct $W$:
\begin{eqnarray} 
N_{ij} \buildrel \rho \circ \over  \longrightarrow E_{ij} + W_{ij}\\
N_{ij} \buildrel \rho \varepsilon \bullet \over  \longrightarrow E_{ij} + W_{ij},
\end{eqnarray}
{and the unique ability of $\sigma_2$ cells to degrade waste: 
\begin{equation} 
W_{ij} \buildrel \rho(1- \varepsilon) \bullet \over  \longrightarrow \emptyset.
\end{equation}
The $\sigma_2$ cells can allocate resources for waste degradation, at the cost of reduced energy production and therefore proliferation, following a linear trade-off ($1\geq\varepsilon\geq0$) consistent with a maximum metabolic load and shared resources for protein synthesis. Therefore, the temporal dynamics of metabolites $N_{ij}$, $E_{ij}$, $W_{ij}$ for a given position $(i,j)$ in the lattice are described by:

\begin{align}
{\partial N_{ij} \over \partial t} = \overbrace{D_{N} \bigtriangledown^{2} N_{ij}}^\text{diffusion term} + \overbrace{\vphantom{\bigtriangledown^{2}}\mu} ^\text{input rate} \nonumber \\ 
- (\underbrace{\vphantom{\delta_{(\sigma_2,S_{ij})}}\eta_{N}}_\text{decay} + \underbrace{\vphantom{\bigtriangledown^{2}}\xi \delta_{(\sigma_1,S_{ij})}}_\text{absorption by $\sigma_1$} + \underbrace{\vphantom{\bigtriangledown^{2}}\varepsilon  \xi \delta_{(\sigma_2,S_{ij})}}_\text{absorption by $\sigma_2$}) N_{ij}
\end{align}
\begin{align}
{\partial E_{ij} \over \partial t} = (\overbrace{\vphantom{\bigtriangledown^{2}}\xi \delta_{(\sigma_1,S_{ij})}}^\text{intake by $\sigma_1$} + \overbrace{\vphantom{\bigtriangledown^{2}}\varepsilon  \xi \delta_{(\sigma_2,S_{ij})}}^\text{intake by $\sigma_2$ }) N_{ij} - \overbrace{\vphantom{\bigtriangledown^{2}}\eta_{E}  E_{ij}}^\text{decay}
\end{align}
\begin{align}
{\partial W_{ij} \over \partial t} = \overbrace{D_{W} \bigtriangledown^{2} W_{ij}}^\text{diffusion term} + (\overbrace{\vphantom{\bigtriangledown^{2}}\xi \delta_{(\sigma_1,S_{ij})}}^\text{$\sigma_1$-produced} + \overbrace{\vphantom{\bigtriangledown^{2}}\varepsilon  \xi \delta_{(\sigma_2,S_{ij})}}^\text{$\sigma_2$-produced}) N_{ij} \nonumber \\
- (\underbrace{\vphantom{\delta_{(\sigma_2,S_{ij})}}\eta_{W}}_\text{decay} + \underbrace{(1 - \varepsilon)  \xi \delta_{(\sigma_2,S_{ij})}}_\text{degradation by $\sigma_2$}) W_{ij}
\end{align}

Here, $\mu$ is the rate of input of the nutrient resource, and $D_{N}$ and $D_{W}$ correspond to the diffusion rate of nutrient and waste respectively. It is worth noting that, being $E$ an intracellular metabolite, it does not diffuse through the lattice. Variables $\eta_N$, $\eta_W$ and $\eta_E$  correspond to the exponential decay parameter of each of the metabolites, while $\xi$ defines the maximal absorption rate. The trade-off parameter $\varepsilon$ adjusts the proportion of nutrient allotted to energy production or waste degradation in $\sigma_2$ cells. Taking into account spatial dynamics of metabolites, both kinds of cells can die either due to local excess of toxic waste or due to lack of internal energy:
\begin{eqnarray} 
\{\circ, \bullet \} \buildrel W_{ij} \ge \Theta_1  \over  \longrightarrow \emptyset\nonumber
\\
\{\circ, \bullet \} \buildrel  E_{ij} \ge \Theta2 \over  \longrightarrow \emptyset
\nonumber
\end{eqnarray}
where $\{\circ, \bullet \}$ indicates that the process equally affects both types of cells. $\Theta_1$ is the maximum value of toxic waste a cell can sustain, $\Theta_2$ is the minimum value of inner energy needed for survival, and $\Theta_3$ defines the inner energy threshold needed for a cell to divide, provided an empty position exists in its neighbourhood:
\begin{equation} 
\{ \circ, \bullet \} \buildrel  E_{ij} \ge \Theta3 \over  \longrightarrow 2 \{ \circ, \bullet \} \nonumber.
\end{equation} 
Mother and daughter cells have the same properties. Energy is equally split between the two cells after division.
\begin{figure}[hbpt]
\label{meanfield}
\begin{center}
\includegraphics[scale=0.27]{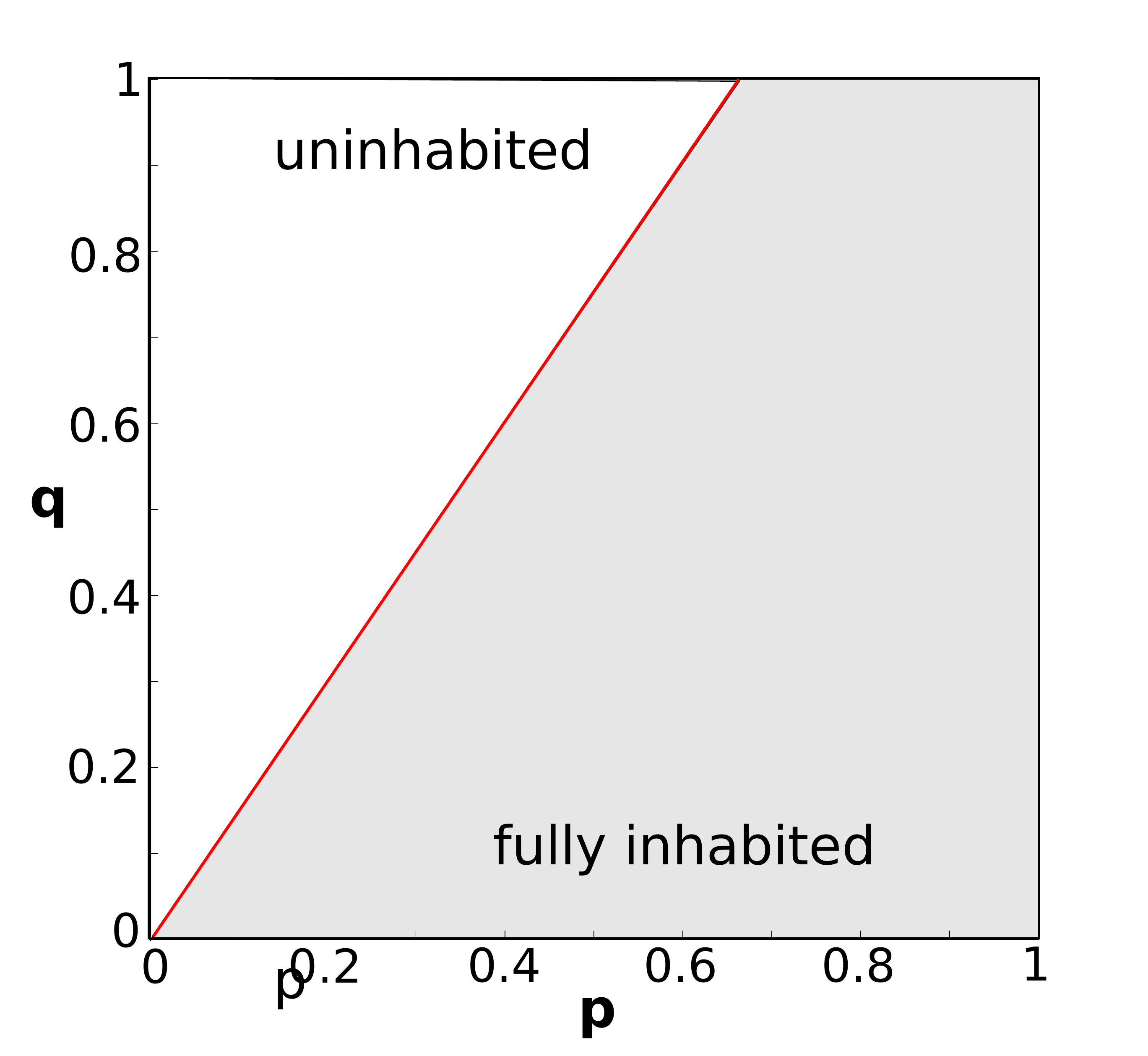}
\end{center}
\caption{Separation between fully inhabited and uninhabited domains in the ($p$,$q$) phase space in the computational mean-field model, for $\varepsilon=0.7$. See Fig. (SM-4) to see how the separation slope varies at different values of $\varepsilon$. }   
\end{figure}
\section{Mean-field approach}

To better understand the general properties of the model, we developed a mean-field ($MF$) approach to the set of ODEs that constitute the metabolism of cells in our system. As a starting point we use the waste differential equation (9). In a well mixed scenario there is no spatial structure and all variables are homogeneous. Hence, the diffusion term and the position subindices cease to be of relevance. Therefore, at the steady state we get: 
\begin{align}
\xi N^* \Bigg[P_{\sigma_1} + \varepsilon P_{\sigma_2} - {(1-\varepsilon) W^* \over N^*} P_{\sigma_2} -{\eta_W W^* \over \xi N^*}\Bigg] = 0 ,
\nonumber
\end{align} 
where $N^*$ and $W^*$ are the equilibrium concentrations of nutrient and waste respectively, and $P_{\sigma_n}$ is the probability that a cell in the system has state $\sigma_n$. Here we separate the analysis into two solutions:
\begin{align}
	\left\{
     		\begin{array}{lr}
       			\xi N^* = 0 \\
			\\
       			W^* = N^*(P_{\sigma_1} + \varepsilon P_{\sigma_2}) / [\eta_W / \xi + (1 - \varepsilon) P_{\sigma_2}]
    		 \end{array}
   	\right.
\end{align} 
$W^*$ is equal to zero when $N^*=0$ (trivial unstable solution) or when the second nullcline is met. 
To develop the mathematical treatment we assume that the two populations of cell states $\sigma_1$ and $\sigma_2$ are at equilibrium. The ratio between populations can be then deduced from the persister cell population dynamics:
$$
{\partial P_{\sigma_1} \over \partial t} = \xi N P_{\sigma_1} + \kappa (q P_{\sigma_2} - p P_{\sigma_1}) - \eta_{\sigma_1} (W, N) P_{\sigma_1}$$
$$ {\partial P_{\sigma_2} \over \partial t}= \xi \varepsilon N P_{\sigma_2} + \kappa (p P_{\sigma_1} - q P_{\sigma_2}) - \eta_{\sigma_2} (W, N) P_{\sigma_2}
$$
The three terms on the right-hand side of these equations represent reproduction, stochastic switching and 
death, respectively. At the steady state, given that: 
$$\xi N P_{\sigma_n} - \eta_{\sigma_n} (W, N) P_{\sigma_n} = 0,$$ 
the relation between the two populations is:
$$P_{\sigma_1} = (q / p) P_{\sigma_2}.$$
Since we have $P_T=P_{\sigma_1}+P_{\sigma_2}$, the expected probabilities for each population at equilibrium are:
\begin{eqnarray}
P_{\sigma_1}  = {q \over (p+q)} \\
P_{\sigma_2}  = {p \over (p+q)}
\end{eqnarray}
Being cell death a threshold function, all cells will die if $W^* > \Theta_1$, and in the opposite case no cell will die. Therefore, whether the population will reach full occupation or not can be determined by incorporating $\Theta_1$ into eq. (1) and transforming the equation into an inequality:
\begin{align}
 	\Theta_1 > {N^* ({q / p} + \varepsilon) \over \Big[1 - \varepsilon +  {\eta_W \over \xi} {(p + q) \over  p} \Big]} 
\end{align} 
This expression defines the region of the parameter space in which, even at maximum population, $W^* < \Theta_1$ and cells do not die. Reordering the terms, we obtain: 
\begin{align}
 	q < p \Bigg[ {\Theta_1 (1 + \eta_N / \xi) / N^* - \varepsilon (1 + \Theta_1/ N^*) \over (1 - \Theta_1 \eta_N / \xi N^* )} \Bigg].
\end{align} 
This inequality defines the boundary dividing the inhabited from the uninhabited region in the ($p,q,\varepsilon$) phase space. In Fig. (SM-4) the boundary for different values of $\varepsilon$ is shown. If waste degradation performed by $\sigma_2$ cells is far greater than the passive decay term ($-\eta_W W$), then the denominator of eq. (13) becomes:
\begin{align}
 	1 - \varepsilon + {\eta_W \over \xi} {(p + q) \over  p} \approx 1 - \varepsilon \nonumber
\end{align} 
hence eq. (12) gets simplified to:
\begin{align}
 	\Theta_1 > {N^* ({q / p} + \varepsilon) \over (1 - \varepsilon )}
\end{align} 
and the inequality in eq. (14) becomes simpler:
\begin{align}
 	q < p \Bigg[ {\Theta_1 \over N^*} - \varepsilon (1+ {\Theta_1 \over N^*}) \Bigg].
\end{align} 
The concentration of nutrient at equilibrium is given by:
\begin{align}
 	N^* ={ \mu \over \eta_N + \xi {(q+\varepsilon p) \over (p+q)}}.
\end{align} 

\section{Results}
\subsection{Mean-field model}
Equations (14) and (16) show that the boundary that in the MF model separates the inhabited from the uninhabited domain depends on $p$, $q$, $\varepsilon$ and on the other parameters of the model appearing in the equations. As $\Theta_1$, $N^*$, $\eta_n$ and $\xi$ have non-zero, positive values, and $p$, $q$ and $\varepsilon$ are constrained to the range $[0, 1]$, it can be shown that this boundary, if represented in the ($p,q$) phase space, has a positive linear slope for $\varepsilon = 0$, which decreases non-linearly  as $\varepsilon$ increases, as shown in Fig. (SM-4). 

We tested this MF prediction against the actual simulations of the EC model, finding also in the latter a boundary, now defining a sharp transition between full occupation and sparse structures. The slope is very similar in the EC and MF models (Fig. 6), indicating that the latter properly captures some features of the EC model.

Equations (14) and (16) also show that there exists a critical $\varepsilon_c$ beyond which the slope is negative: 
\begin{displaymath}
\varepsilon_c =
	\left\{
     		\begin{array}{lr}
       			(1 + \eta_N / \xi) / (1 + N^* /\Theta_1) \quad $(from eq. 14)$ \\
			\\
       			1 / (1 + N^* /\Theta_1) \quad \quad \quad \quad \quad $(from eq. 16)$
    		 \end{array}
   	\right.
\end{displaymath}

Under such conditions no pair $p, q$ in $[0, 1]$ can make the inequality true, resulting in a system dominated by death processes. In the simplified scenario described by inequality (16), $\varepsilon_c$ is always in the range $[0, 1]$, but if we consider the non-simplified eq. (14), then $\varepsilon_c$ can assume a value out of the $[0, 1]$ range, when $\eta_N \Theta_1 > \xi N^*$. 

\subsection{Pattern formation in the DA-SPS model}
\begin{figure}[h]
\label{patterns}
\begin{center}
\includegraphics[scale=0.64]{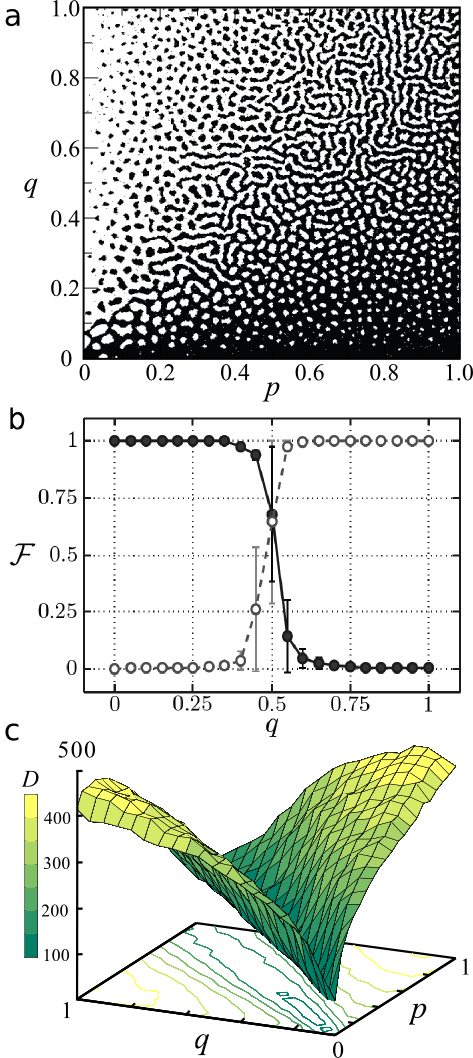}
\end{center}
\caption{
(a) Pattern formation on a fully occupied lattice in which the individual transition probabilities vary spatially from $0$ to $1$, $\kappa = 10 ^{-3}$, $T = 10$, with a symmetric adhesion matrix and after $10^5$ iterations. Different arrangements -from spots to stripes to mazes- of both cell types are attained, reminiscent of fixed wavelength structures. (b) Average reachable fraction (${\cal F}$) of cells of a particular type (white and black circles represent $\sigma_1$ and $\sigma_2$ cells respectively), fixed $p=0.5$. (c) Average domain count ($D$) -defined as groups of contiguous cells with the same state- for each pair ($p$, $q$).}   
\end{figure}
\begin{figure*}[htb]
\label{kappa}
\begin{center}
\includegraphics[scale=0.166]{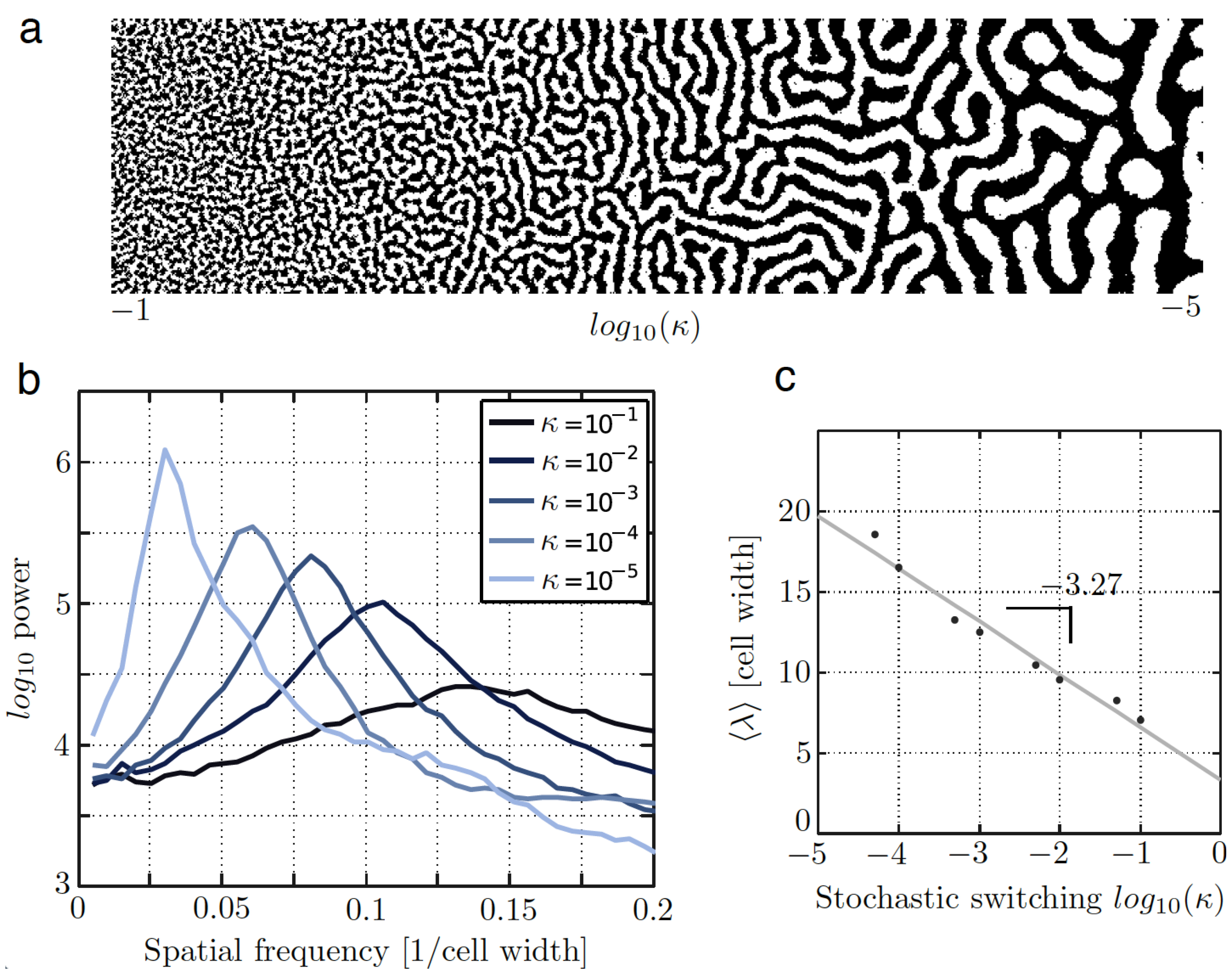}
\end{center}
\caption{
(a) The effect of changing the coupling factor between adhesion processes and stochastic phenotype switching. Here $p=q=0.5$, and $log_{10} (\kappa)$ is tuned continuously from $-1$ to $-5$. (b) Dominant wavelength for the same range of $\kappa$ values using a radially averaged power spectrum. For discrete values of $\kappa$ (in hues of blue) we show the existence of a moving peak in the frequency domain, with decreasing frequencies -i.e. with increasing wavelength- as $\kappa$ decreases. (5 replicates of a $250\times250$ lattice, $t=10^6$ algorithm iterations, processed with Matlab 2013b FFT). (c) Average value of the dominant wavelength plotted against $log_{10}(\kappa)$. We also show a possible fitting of the data.}  
\end{figure*}

In the DA-SPS expansion of Potts model, cells can perform phenotypic switching through a transition rule regulated by transition probabilities $p$ and $q$, and their movement is driven by differential adhesion and the tendency to minimize interaction energy between cells. Parameter $\kappa$ is introduced as a scaling factor which regulates the relative speed at which cell sorting by adhesion and phenotypic switching occur.
Using this model, we assess the role in pattern formation of the individual transition probabilities ($p$, $q$) for a fixed $\kappa$. This analysis reveals that, in spite of model simplicity, $\sigma_1$ and $\sigma_2$ cells are able to self-organize in space in periodic structures. These patterns can range from spots to stripes to mazes, depending on the relative values of $p$ and $q$ (Fig. 4a).

To characterize the phases of the morphospace we applied a standard percolation algorithm to the final macro-state of each simulation. Figure (4b) shows that the average reachable fraction of each of the two states $\sigma_1$ and $\sigma_2$ displays a sharp transition. This defines three clear regimes: non-percolating (spots), percolating (fully connected maze) and transition regime, where spots become stripes of increasing length and are marginally able to extend to other domains. 
Interestingly, this transition occurs for a different value of the control parameter with respect to non-correlated percolation studies \cite{Malarz2005} ($q=0.5$ instead of $q_c \approx 0.592$).
For this value the structures of both cell states percolate, giving rise to the labyrinth phase. Although percolation analysis shows a sharp transition, the number of domains -i.e. clusters of lattice sites with same state- varies smoothly over the phase space (Fig. 4c).
Another interesting feature that can be observed in Fig. (4a, 4c) is that even if the ratio $p / q$ remains constant, lower values of the transition parameters generate bigger structures with fewer domains. 

Since in our model $\kappa$ is a scaling factor for both $p$ and $q$, we set out to quantify the effect of this parameter in the pattern formation process. Fig. (5a) shows the qualitative impact of the tuning of $\kappa$ in a full-occupation DA-SPS model. At low $log_{10}(\kappa)$ values, SPS occurs at a slower pace than the cell sorting process, which is therefore able to properly separate cells in two phases. Instead, for high $log_{10}(\kappa)$ values, SPS occurs at a faster pace than the cell sorting process, which brings about an almost random distribution of states. 

\begin{figure*}[hbtp]
\label{pqvary}
\begin{center}
\includegraphics[scale=0.44]{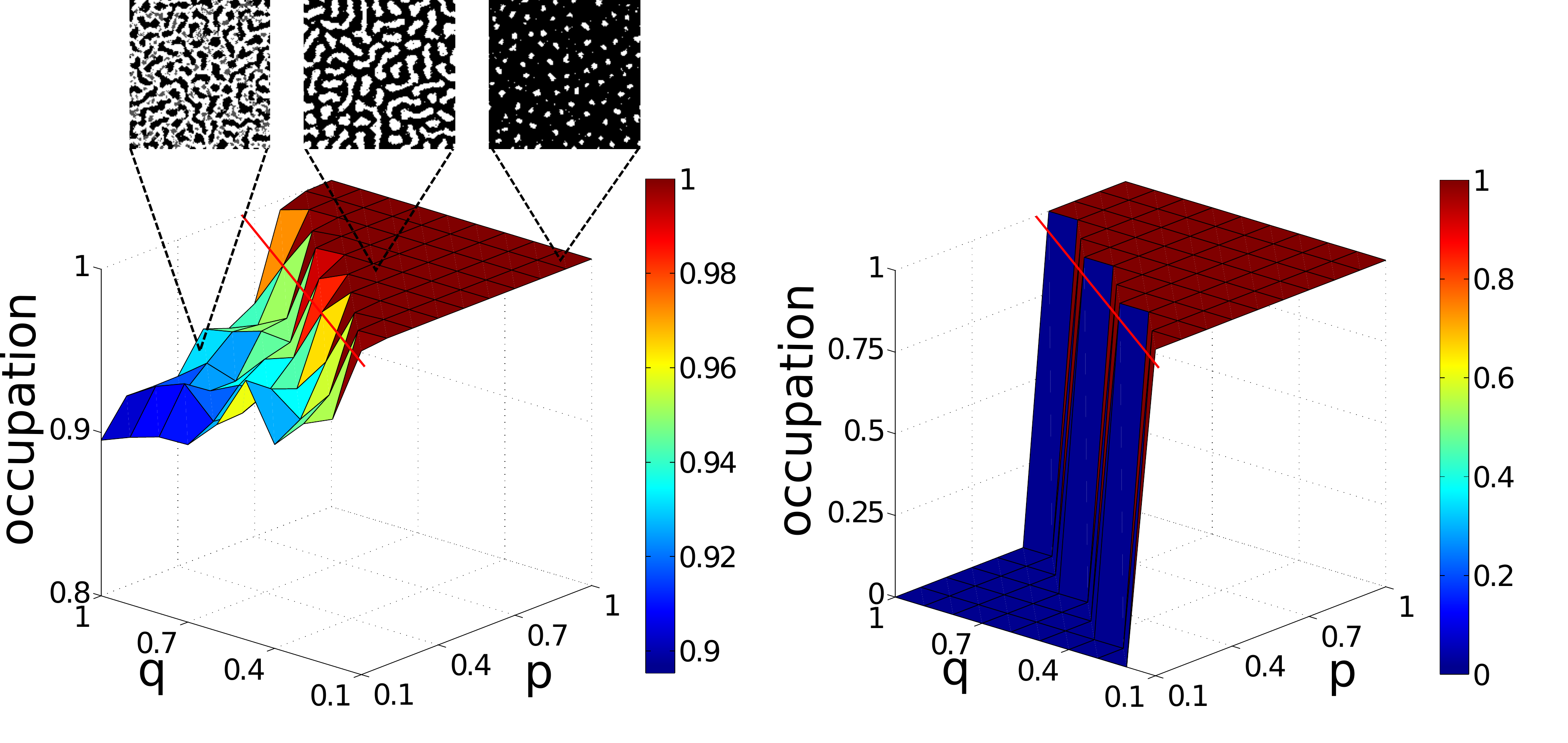}
\end{center}
\caption{Occupation levels in the \emph{Ecology and competition} (EC) model (left) and \emph{mean-field} (MF) simulations (right), with $\varepsilon=0.7$ and $p$ and $q$ varying from 0.1 to 1. The occupation level for each pair of values ($p,q$) is obtained from a separate simulation in a $150\times150$ thoroidal lattice. The red line in both images divides the fully inhabited region from the uninhabited or partially inhabited region of the ($p,q$) phase space. Mean-field simulation  reproduces the MF analytical model results, as expected. In the EC model simulation, only few cells die in the partially inhabited region and occupation level remains high because the spatial inhomogenity of metabolites allows cells (for the chosen parameters) to survive and overcome the presence of peaks of toxic waste. The boundary at which the transition happens changes depending on the value of $\varepsilon$, as can be shown by the analytical MF model, the slope becoming smaller at higher values of $\varepsilon$ (see Fig. SM-4). As an example we show the patterns of some lattices for given values of $p$ and $q$ the EC model (white for $\sigma_1$ cells, black for $\sigma_2$ cells, gray for medium). An alternative representation of the same result is shown in Fig. (SM-2).}   
\end{figure*}

Following the same strategy as before, we ran a set of simulations with periodic boundary conditions, transition probabilities $p=q=0.5$, and different values of $\kappa$. We then applied a standard Fourier Transform analysis in order to confirm the existence of dominant frequencies in the spatial distribution of lattice states after $10^6$ iterations. Fig. (5b) displays the spatial frequency contribution in a radially averaged power spectrum (RAPS). The presence of a single peak in the RAPS indicates that the periodic structures are built by a single dominant frequency without any specific orientation in the spatial domain, i.e. we obtain fixed wavelength structures reminiscent of Turing patterns. Furthermore, the peak position and width are subject to the particular value of $\kappa$, specifically the wavelength of the pattern decreases as $log_{10}(\kappa)$ increases. The particular mathematical relation between these two variables is displayed in Fig. (5c).

\subsection{\label{sec:level2}Phase transitions and fitness landscapes}

\begin{figure*}[hpbt]
\label{adhstruct}
\begin{center}
\includegraphics[scale=0.5]{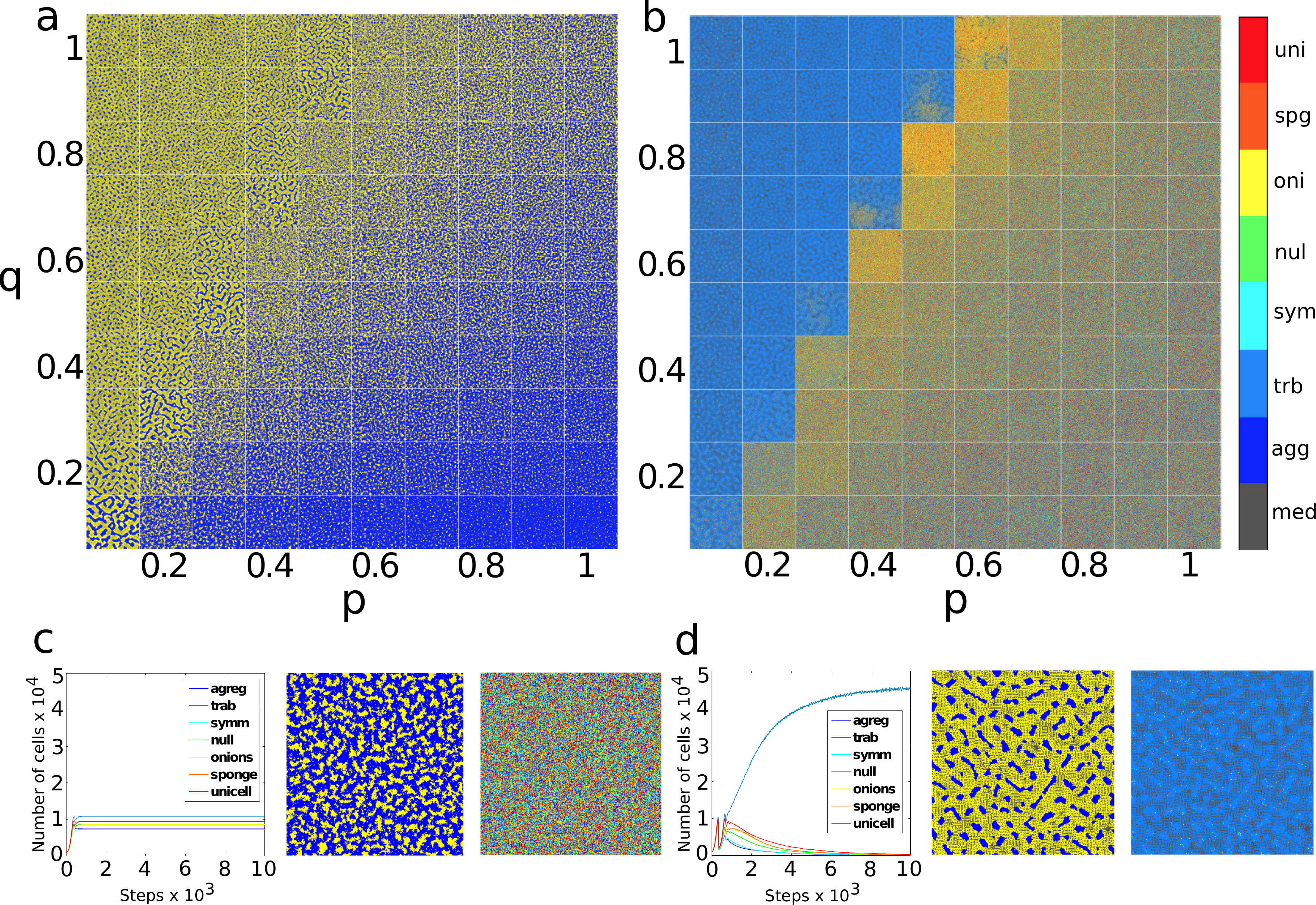}

\end{center}
\caption{Collage of 100 runs at fixed $\varepsilon=0.7$ and varying $p$ and $q$, where populations of cells differing only by their adhesion matrices are put in competition, to assess which adhesion matrix allows the corresponding population to gain higher fitness. In a) we represent the disposition in the lattice at the end of the simulation for the two cell phenotypes $\sigma_1$ (white cells, represented in yellow), and $\sigma_2$ (black cells, represented in blue), regardless of what cell population they belong to. In b) we represent the disposition in the lattice at the end of the simulation for cells of each population, represented in different colors. In the fully inhabited region below the boundary, no death occurs, the populations grow at similar rates and saturate the lattice in similar proportions. In the region of the phase space above the boundary, cell death occurs, allowing the different cell populations to compete for lattice occupation. In this region, the \emph{trabecular} population shows the highest fitness over other populations. The two phases are delimited by the slope separating the empty from the full occupation region in the mean-field analytical model. In c) and d) we represent the change of occupation level in time for each population, at ($p=1, q=0.6$) and at ($p=0.2, q=0.6$) respectively. See Fig. (SM-3) for the occupation levels of each population at varying $p$ and $q$.}   

\end{figure*}

In the \emph{ecology and competition} (EC) model cells are not only subject to adhesion processes and phenotypic switching, but can absorb the substrate which is constantly produced all over the lattice, and transform it into energy only ($\sigma_1$ cells) or energy and waste ($\sigma_2$ cells).
We compare the EC model with the \emph{mean-field} (MF) model}, to understand which properties can be predicted from the latter and which ones instead emerge from the complexity introduced by spatial organization and inhomogeneities in the levels of waste and substrate.  
We run a set of simulations, each with a different pair of $p$ and $q$ values, using periodic boundary conditions and fixing $\varepsilon=0.7$ and $k=10^{-3}$. The $\kappa$ parameter is fixed at $10^{-3}$, in such a way that adhesion processes occur $10^3$ times faster than phenotypic switching. In the initial time step, $100\%$ of lattice is occupied by randomly distributed  $\sigma_1$ and $\sigma_2$ cells, being the ratio between the two phenotypes already set at the equilibrium value following (Eq. $11$, $12$), depending on $p$ and $q$ values.
The results in Fig. (6) and in Fig. (SM-2) show a clear correspondence between the analytical MF result and the simulated MF, indicating that the model we implemented properly reproduces the expected theoretical results. In the EC model, cells can also occupy the region of the phase space which was left empty in the MF model. In fact, while in the MF model simulation all cells die instantaneously when the level of waste reaches $\Theta_1$ (the maximum level of $W$ a cell can sustain), in the EC model the death of a few cells can reduce the pressure on the system and allow population survival also in those parametrically unfavoured habitats where waste concentration can locally exceed $\Theta_1$. 

The slope in the ($p,q$) phase space separating inhabited from uninhabited region in the MF model varies depending on the value of $\varepsilon$ and of other relevant parameters of the model (Fig. SM-4). For $\varepsilon=0$, black and white cells are indistinguishable, and their behaviour is independent from the value of $p$ and $q$, as can be calculated in the MF analytical model. As $\varepsilon$ increases, $\sigma_2$ cells can degrade waste better, but are also less able to elaborate nutrient and can die from lack of energy, unless $p$ and $q$ are set in such a way that there is a high probability for the $\sigma_2$ cell to switch to a $\sigma_1$ cell before inner $E$ value gets below $\Theta_2$ (the minimum value of $E$ a cell needs for survival).

At steady state different structures emerge, depending on $p$, $q$ and $\varepsilon$ values. The results for $\varepsilon=0.7$ described in this paragraph are shown in Fig. (6) and in Fig. (SM-2). For high values of $p$ and $q$, $\sigma_1$ and $\sigma_2$ cells form Turing Patterns in a percolating structure. For high values of $p$ and low values of $q$, $\sigma_2$ cells dominate (note that for $\varepsilon=0.7$ they are still able to degrade W so as not to reach $\Theta_1$). For lower values of $p$ and higher values of $q$, $\sigma_1$ cells substitute $\sigma_2$ cells less rapidly, $\Theta_1$ can be reached more easily and cells start to die. 

Lastly, we want to assess whether our EC model, integrating DA, SPS and metabolism in a habitat with nutrient and toxic waste, can present situations in which structural organization can affect the fitness of cells in the habitat. To do so, we launch a set of simulations in which cells with different adhesion matrices compete for reproduction. At the beginning of the simulation $10\%$ of the lattice is occupied by cells, which are divided in equal populations differing only by type of adhesion matrix, which can be \emph{aggregate}, \emph{trabecular}, \emph{symmetric}, \emph{null}, \emph{onions}, \emph{sponge} or \emph{unicellular} (see Fig. SM-1). The adhesion matrix type is transmitted by each cell to its offspring. Cells are randomly distributed over the lattice regardless of the population they belong to. It is important to stress that in terms of preferential attachment, cells only sense the states $\sigma_1$ and $\sigma_2$ of neighbouring cells, independently from population type.

For each set of parameters $\varepsilon$, $p$ and $q$ in the phase space, we assess which type of adhesion matrix brings the related population to maximum fitness, by comparing the level of occupation of the lattice for each population. In Fig. (7a) we represent cells of state $\sigma_1$ or $\sigma_2$ independently from their population, while in Fig. (7b) we show the same cells differentiated by population. At the bottom (Fig. 7c, 7d) we represent the change of occupation level in time for each population. In the region of the phase space which is fully inhabited in the MF model, cells can survive independently from their adhesion matrix values and never die. For this reason the populations related to the various adhesion matrices are equally numbered in this area -with slight differences due only to growth speed before the lattice becomes saturated- and randomly distributed over space, with no emerging structure. However, in the domain of the phase space which was uninhabited in the MF, $p$ and $q$ have such values that do not guarantee survival of cells. Since cell death may occur in the EC model for this region of the phase space, here the values of the adhesion matrix do make a difference and some species get selected over others. In particular, the population with `trabecular' adhesion matrix prevails. Moreover, we can observe that in this area cells organize in a maze structure, exhibiting division of labour within the same population. Lastly, in the frontier between the two zones various populations can coexist, with a prevalence for `onion' adhesion matrix at high values of $q$. In Fig. (SM-3) we show the relative occupation levels of each population at varying $p$ and $q$.

\section{Discussion and prospects}



In this paper we have shown a novel way of constructing periodic arrangements of cell types in the form of a hybrid differential adhesion and stochastic switching process. This mechanism does not rely on differential diffusion (normally found between the activator and inhibitor species in canonical Turing-type systems) yet it can create the same kind of structures in a predictable, scalable way. The key ingredients proposed at this level include the differential adhesion hypotheses stemming from Steinberg's work (that considers the minimization dynamics associated with a set of interacting spins or adhesion strengths) and genetic switches following Markovian stochastic dynamics, which are the source of cell diversity and the basis of some adaptive responses displayed by microbial populations \cite{Lewis2010}. The switching dynamics can modify the types of patterns expected from the purely energy-driven scenario, thus indicating that potential forms of phenotypic change can lead to additional richness of pattern forming rules. A range of spatially ordered structures is obtained displaying characteristic length scales. Being both key ingredients present in extant organisms, we consider that this simple mechanism might have been used originally (and might be reproduced in the future by synthetic means) to create regular structures in aggregates and colonies. 

In relation to pattern formation dynamics, our hybrid adhesion model offers an alternative way to generate Turing Patterns, which were up to now directly related with Turing's RD mechanisms mediated by a diffusible molecule, or with apparently unrelated but mathematically equivalent systems such as  direct contact-mediated regulation by means of which cells are affecting each other's internal rates of reactions \cite{Babloyantz1977}. In fact, differently from what was proposed by Babloyantz, in our hybrid DA-SPS model the molecules on the surface of one cell do not affect the rates of reactions in its neighbours: the phenotypic switching process occurs in any cell independently from its past and from its neighbour's state, and it is not influenced by the values of the adhesion matrix. 

The second relevant aspect considered is how these forming structures might be of benefit to a developing cooperative population in presence of nutrient resources and toxic agents. To do so we developed an ecology and competition model where a minimal metabolism enables positive or negative interactions between cells. Cells can cooperate by metabolizing waste byproducts, yet they will suffer from decreased growth rates at higher population densities due to substrate attrition. In the EC model further pattern-forming processes can be predicted.

To further asses how structural organization can affect the fitness of cells in the habitat, we studied how cells with different adhesion matrices compete for reproduction. 
Interestingly when many populations differing in terms of adhesion properties compete, in the region of the phase space with strong selective pressures only one of the populations survives (Fig. 7). The selected specie consistently develops a periodic multicellular structure which is superior to both the unicellular and the unstructured multicellular one, suggesting that higher order properties might be of relevance to the establishment of functionality and cooperation. 

This simple competition model shows how minimal interaction properties pervading the metabolism of multiple species might come to play a central role in forcing the transition to collective fitness and behaviour, and sets the groundwork for explicitly evolutionary automata \cite{Duran2015}, where cells can optimize several genotype dimensions in order to attain more resources.

\section{Acknowledgements}
We thank the members of the Complex Systems Lab for useful discussions, and Amad\'is Pag\`es for useful hints on code debugging. This work has been supported by the Bot\'in Foundation by Banco Santander through its Santander Universities Global Division, a MINECO fellowship and by the Santa Fe Institute.


\begin{thebibliography}{72}

\bibitem{Bonner1988}
Bonner, J. T. 1998. {\em The evolution of complexity by means of natural selection}. 
Princeton University Press. Princeton.

\bibitem{Carroll2001}
Carroll, S.B. 2001. {\em Chance and necessity: the evolution of morphological complexity and diversity}. Nature 409,1102-1109
\bibitem{Nedelcu2015}
Nedelcu, A.M. and Ruiz-Trillo, I. 2015 (eds.) {\em Evolutionary Transitions to Multicellular Life: Principles and Mechanisms}. Springer-Verlag, London.

\bibitem{Bonner2001}
Bonner, J. T. 2001. {\em First signals: the evolution of multicellular development}. 
Princeton University Press. Princeton.

\bibitem{Newman1990}
Newman, S. and Comper, W. D. 1990. {\em Generic physical mechanisms of morphogenesis and pattern formation}. Development 110, 1-18.

\bibitem{Waddington}
Waddington, C. A. 1956. {\em Principles of Embriology}. George Allen Ltd. London.

\bibitem{Turing1952}
Turing, Alan. 1952. {\em The chemical basis of morphogenesis}. Phil. Trans. Royal Soc. London B 237,  37-72.

\bibitem{Rashevsky}
Rashevsky, N. 1948. {\em On periodicities in metabolizing systems. The Bulletin of mathematical biophysics}. 10(3), 159-174.

\bibitem{Murray1981}
Murray, J. D. 1981. {\em A pre-pattern formation mechanism for animal coat markings}. Journal of Theoretical Biology, 88(1), 161-199.

\bibitem{Sole1}
Sol\'e R., Bascompte J. and Valls J. 1993. {\em Stability and Complexity of Spatially Extended Two-species Competition}. J. Theor. Biol. 
159, 469-480.

\bibitem{Sole2}
Bascompte J. and Sol\'e R. 1995.  {\em Rethinking Complexity: Modelling Spatiotemporal Dynamics in Ecology}. Trends Ecol. Evol. 10, 361-366.

\bibitem{TheraulazPNAS}
Theraulaz, G., Bonabeau, E. Nicolis, S. C., Sol\'e, R. et al 2002. {\em Spatial patterns in ant colonies}. {\em Proc. Natl. Acad. Sci USA.} 99: 9645-9649.

\bibitem{Jilkine2011}
Jilkine A. and Edelstein-Keshet, L. 2011. {\em A Comparison of Mathematical Models for Polarization of Single Eukaryotic Cells in Response to Guided Cues}. 
PLoS Comput. Biol. 7, e1001121.

\bibitem{Meinhardt2000}
Meinhardt, H., and Gierer, A. 2000. {\em Pattern formation by local self-activation and lateral inhibition}. Bioessays, 22(8), 753-760.

\bibitem{Goodwin1994}
Goodwin, B. C. 1994. {\em How the leopard got its spots}. Princeton U. Press, Princeton.

\bibitem{koch}
Koch, A.J. and Meinhardt, H. 1994. {\em Biological Pattern Formation : from Basic Mechanisms to Complex Structures} Rev. Modern Physics 66, 1481-1507.

\bibitem{lejeune}
Lejeune, O., and Tlidi, M. 1999. {\em A model for the explanation of vegetation stripes (tiger bush)}. Journal of Vegetation science, 10(2), 201-208.

\bibitem{sick}
Sick, S., Reinker, S., Timmer, J., and Schlake, T. 2006. {\em WNT and DKK determine hair follicle spacing through a reaction-diffusion mechanism}. Science, 314(5804), 1447-1450.

\bibitem{econoumou}
Economou, A. D., Ohazama, A., Porntaveetus, T., Sharpe, P. T., Kondo, S., Basson, M. A. and Green, J. B. 2012. {\em Periodic stripe formation by a Turing mechanism operating at growth zones in the mammalian palate}. Nature genetics, 44(3), 348-351.

\bibitem{raspopovic}
Raspopovic, J., Marcon, L., Russo, L., and Sharpe, J. 2014. {\em Digit patterning is controlled by a Bmp-Sox9-Wnt Turing network modulated by morphogen gradients}. Science, 345(6196), 566-570.

 \bibitem{Goel1970}
Goel, N., Campbell, R. D., Gordon, R., Rosen, R., Martinez, H., and Ycas, M. 1970. {\em Self-Sorting of Isotropic Cells}. 
 {\em J. Theor. Biol.} 28 (3): 423-68. 

\bibitem{NewmanBaht2008}
Newman, S. A. and Baht, R. 2008. {\em Dynamical patterning modules: physico-genetic determinants of morphological development and evolution}.  {\em Phys. Biol.} 5: 015008.

\bibitem{ForgacsNewman2005}
Forgacs, G., and Newman, S. A. 2005. {\em Biological physics of the developing embryo}. Cambridge U. Press, Cambridge.

\bibitem{ForgacsFoty1998}
Forgacs, Gabor, et al. {\em Viscoelastic properties of living embryonic tissues: a quantitative study}. Biophysical journal 74.5 (1998): 2227-2234.

\bibitem{Foty2005}
Foty, R. A., and Steinberg, M. S. 2005. {\em The differential adhesion hypothesis: a direct evaluation}. 
 {\em Dev. Biol.} 278(1): 255-263.
 
\bibitem{NewmanBaht2009}
Newman, Stuart A., and Ramray Bhat. {\em Dynamical patterning modules: a" pattern language" for development and evolution of multicellular form}. International Journal of Developmental Biology 53.5 (2009): 693.
 
 \bibitem{Steinberg1975}
Steinberg, M S. 1975. {\em Adhesion-Guided Multicellular Assembly: A Commentary upon the Postulates, Real and Imagined, of the Differential Adhesion Hypothesis, with Special Attention to Computer Simulations of Cell Sorting}. 
 {\em J. Theor. Biol.} 55 (2): 431-43.
 
 \bibitem{GranerGlazier1992}
Graner, F., and J. A. Glazier. {\em Simulation of biological cell sorting using a two-dimensional extended Potts model}. Physical review letters 69.13 (1992): 2013.

\bibitem{Hogeweg2000}
Hogeweg, P. 2000. {\em Evolving Mechanisms of Morphogenesis: 
on the Interplay between Differential Adhesion and Cell Differentiation}. 
{\em J. Theor. Biol.} 203: 317-333 

\bibitem{Hallet2001}
Hallet, B. 2001. {\em Playing Dr Jekyll and Mr Hyde: combined mechanisms of phase variation in bacteria}. 
{\em Current opinion in microbiology}. 4(5), 570-581.

\bibitem{Balaban2004}
Balaban NQ, Merrin J, Chait R, Kowalik L and Leibler S. 2004. {\em Bacterial persistence as a phenotypic switch}.
 {\em Science} 305:1622-1625.

\bibitem{Darmon2014}
Darmon, E., and D.R.F. Leach. 2014. {\em Bacterial Genome Instability}. {\em Microbiology and Molecular Biology Reviews} : MMBR 78 (1): 1-39.

\bibitem{Malarz2005}
Malarz, Krzysztof and Galam, Serge 2005. {\em Square-lattice site percolation at increasing ranges of neighbor bonds}.
 {\em Phys. Rev. E} 71

 \bibitem{Lewis2010}
Lewis K 2010. Persister cells.  { \em Annu. Rev. Microbiol}. 64: 357-372.

\bibitem{Babloyantz1977}
Babloyantz, Agnessa. 
 {\em Molecules, dynamics, and life: An introduction to self-organization of matter.} Wiley-Interscience, 1986.

\bibitem{Duran2015}
Duran-Nebreda, S., Bonforti, A., Monta\~{n}ez, R., Valverde, S., and Sol\'e , R. 2015. {\em Emergence of proto-organisms from bistable stochastic differentiation and adhesion}. {\em arXiv preprint} arXiv:1511.02079.


\end{thebibliography}
\end{document}